\documentclass{acm_proc_article-sp}
\usepackage{listings}
\usepackage{url}

\begin{document}

\title{Design of an Audio Interface for Patmos}
\subtitle{ 02211 Advanced Computer Architecture }
%
% You need the command \numberofauthors to handle the 'placement
% and alignment' of the authors beneath the title.
%
% For aesthetic reasons, we recommend 'three authors at a time'
% i.e. three 'name/affiliation blocks' be placed beneath the title.
%
% NOTE: You are NOT restricted in how many 'rows' of
% "name/affiliations" may appear. We just ask that you restrict
% the number of 'columns' to three.
%
% Because of the available 'opening page real-estate'
% we ask you to refrain from putting more than six authors
% (two rows with three columns) beneath the article title.
% More than six makes the first-page appear very cluttered indeed.
%
% Use the \alignauthor commands to handle the names
% and affiliations for an 'aesthetic maximum' of six authors.
% Add names, affiliations, addresses for
% the seventh etc. author(s) as the argument for the
% \additionalauthors command.
% These 'additional authors' will be output/set for you
% without further effort on your part as the last section in
% the body of your article BEFORE References or any Appendices.

\numberofauthors{2} %  in this sample file, there are a *total*
% of EIGHT authors. SIX appear on the 'first-page' (for formatting
% reasons) and the remaining two appear in the \additionalauthors section.
%
\author{
% You can go ahead and credit any number of authors here,
% e.g. one 'row of three' or two rows (consisting of one row of three
% and a second row of one, two or three).
%
% The command \alignauthor (no curly braces needed) should
% precede each author name, affiliation/snail-mail address and
% e-mail address. Additionally, tag each line of
% affiliation/address with \affaddr, and tag the
% e-mail address with \email.
%
% 1st. author
\alignauthor
Daniel Sanz Ausin\\
       \affaddr{Department of Applied Mathematics and Computer Science}\\
       \affaddr{Technical University of Denmark}\\
       \email{s142290@student.dtu.dk}
% 2nd. author
\alignauthor Fabian Goerge \\
       \affaddr{Department of Applied Mathematics and Computer Science}\\
       \affaddr{Technical University of Denmark}\\
       \email{s150957@student.dtu.dk}
}
% There's nothing stopping you putting the seventh, eighth, etc.
% author on the opening page (as the 'third row') but we ask,
% for aesthetic reasons that you place these 'additional authors'
\date{\today}
% Just remember to make sure that the TOTAL number of authors
% is the number that will appear on the first page PLUS the
% number that will appear in the \additionalauthors section.

%%%------------MAX 8 pages------------------

\maketitle
\begin{abstract}

This paper describes the design and implementation of an audio interface for the Patmos processor, which runs on an Altera \texttt{DE2-115} FPGA board. This board has an audio codec included, the \texttt{WM8731}. The interface described in this work allows to receive and send audio from and to the \texttt{WM8731}, and to synthesize, store or manipulate audio signals writing C programs for Patmos.

The structure of this project is integrated with the Patmos project: new hardware modules have been added as IOs, which allow the communication between the processor and the audio codec. These modules include a clock generator for the audio chip, ADC and DAC modules for the audio conversion from analog to digital and vice versa, and an ${I}^{2}C$ module which allows setting configuration parameters on the audio codec. Moreover, a top module has been created, which connects all the modules previously mentioned between them, to Patmos and to the \texttt{WM8731}, using the external pins of the FPGA.

Simulations have been done using the Patmos emulator to check the correct functionality of all the modules. After that, the Patmos processor with the audio interface has been synthesized and loaded into the Altera FPGA. C programs have been written, and SignalTap, the logic analyzer from Quartus, has been used to verify that the modules work as expected.

The results show that the audio interface works as desired: the implemented C programs show that the audio is input and output to and from Patmos correctly, with the right format and the expected sampling frequency. The analogue audio signal output from the FPGA has been recorded using an external audio interface, and the signals look exactly as anticipated. 

\end{abstract}
 
%http://www.acm.org/sigs/publications/sigguide-v2.2sp
%\terms{Design, Measurement}

%\keywords{Real-time systems, Time-predictable computer architecture, Patmos, T-CREST, DE2-115, WM8731}

\newpage
\section{Introduction}

The audio interface described in this paper is intended to be used with the Patmos processor. Patmos is an open source RISC ISAs with a load-store architecture, that is optimized for Real-Time Systems. Patmos is part of a project founded by the European Union called T-CREST (Time-predictable Multi-Core Architecture for Embedded Systems).\cite{PatmosSpec}

The majority of the hardware development for Patmos was done with the hardware description language Chisel. Chisel is an open-source Scala-based hardware description language developed by the university of Berkeley. \cite{ChiselTut} All the development in this project was also Chisel-based. 

The development is based on the \texttt{DE2-115} FPGA board from Altera. Hence focuses this paper also on the communication with the on board \texttt{WM8731} audio codec from Wolfson.

The paper is organized as follows: The following Section~\ref{sec:RelatedWork} discusses the related work. Section~\ref{sec:Design} introduces the design of the audio interface. Section~\ref{sec:Implementation} covers the implementations in Chisel, VHDL and C as well as the FPGA integration process. After that Section~\ref{sec:Evaluation} presents the evaluation and finally, section~\ref{sec:conclusion} concludes the paper.

\section{Related Work}
\label{sec:RelatedWork}

As mentioned before, this project has been developped using Patmos, the open source real-time RISC processor developped by T-CREST. There are many different versions of the Patmos processor, including a multi-core version with 9 cores. But, for this project, the most simple 1-core version has been used.

%%%This is not realy good, because the reference is of poor quality.
Jos\'{e} Albiach discusses in his master thesis the FPGA implementation for the \texttt{WM8731} audio codec on an Altera UP3 board. In his work he briefly summarizes the functionality and then describes his approach to interact with the audio chip. In his work he suggest to divides the clock to generate the needed clock signals for the audio chip. \cite{thesis}

\section{Design}
\label{sec:Design}

In the Altera data sheet for the \texttt{DE2-115} Board the basic features and concept of the Wolfson \texttt{WM8731} audio encoder and decoder can be found. The \texttt{WM8731} has line-in and line-out ports as well as a microphone-in. The chip can be configured under use of the ${I}^{2}C$ bus to set parameters such as the volume, the sampling frequency, and so on. The FPGA pins connected to the audio chip are for input audio data (\texttt{ADCDAT}) and output audio data (\texttt{DACDAT}) as well as there respective sampling clocks (\texttt{ADCLRCK, DACLRCK}), which are used for synchronization. Furthermore a master clock input for the chip is available (\texttt{XCK}), and an additional clock used for the ADC and DAC (\texttt{BCLK}). Finally, there are two pins for the  ${I}^{2}C$ interface:  data (\texttt{SDAT}) and clock (\texttt{SCLK}). \cite[chapter 4.11]{DE2115DataSheet}

\begin{figure}[ht!]
\centering
\includegraphics[width=0.47\textwidth]{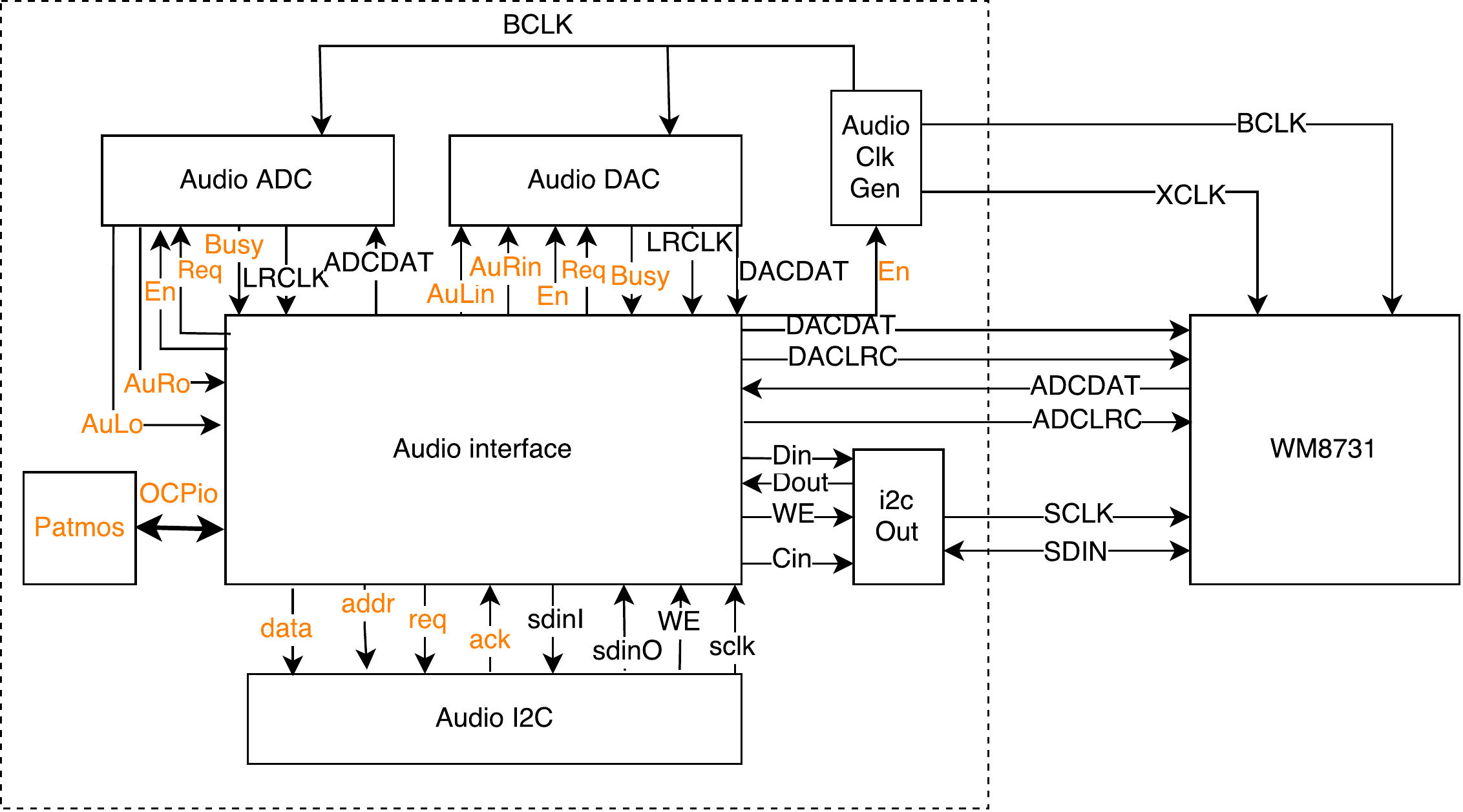}
\caption{Overview of the Design}
\label{fig:DesginOverview}
\end{figure}

It was necessary to consult the data sheet of the \texttt{WM8731} for a better understating. It can be clearly seen from there that there are two main separated parts: the digital audio interface and the control interface. \cite{WM8731DataSheet} According to this, the design was split into smaller modules, in order to keep a good structure of the system. An overview of the design can be seen in Figure~\ref{fig:DesginOverview}. Patmos and the created audio interface communicate via the open core protocol. Applications on the Patmos processor can then write or read audio data and configuration parameters to or from registers in the audio interface. The "AudioDAC" and "AudioADC" components exchange the content of the audio registers with the audio chip in the correct format for the audio interface of the \texttt{WM8731}. Likewise, the "Audio ${I}^{2}C$" component writes the data and address of the configuration registers using the ${I}^{2}C$ protocol. For both the conversion from digital to analog and from analog to digital audio data, sampling clocks are needed. Therefore, the clock generation unit generates the master clock input (\texttt{XCK}) and the digital audio bit clock (\texttt{BCLK}). Later these units will be described in more detail.

\subsection{Audio Interface Component}

The audio interface is designed to be the heart of the implementation which includes all the described components (see Figure~\ref{fig:DesginOverview}). 

The audio interface has different registers in which Patmos can write and read from. Depending on its functionality will some registers only be readable. 

Additionally, this component will specify the desired output pins, which will be 1-bit outputs for the digital to analog data (\texttt{dacDat}) and the sample rate clocks (\texttt{dacLrc}, \texttt{adcLrc}) and a 1-bit input for the analog to digital data (\texttt{adcDat}). The audio chip will be used in slave mode and will be clocked by the FPGA hence there are also pins needed for the master clock (\texttt{XCK})  and the bit clock (\texttt{BCLK}). Moreover, does the ${I}^{2}C$ interface need two outputs (Data abd Clock). The data signal is bidirectional (as it will be described in the next subsection), so it needs both an input and output pin as well as a write enable signal. This 3-state buffer design is used because it is impossible to define \texttt{inout} types in Chisel. 

Furthermore, this unit forwards the default configurations from the hardware configuration XML file to the submodules. The configuration parameters for the audio interface include the audio data length, which can be either 16, 20, 24 or 32-bit, the sampling frequency (Fs) divider, which can range from $8 kHz$ to $96 kHz$, and finally the divider for the clock signals into the audio codec (\texttt{BCLK} and \texttt{XCLK}). For this last signal the divider value chosen is 6, to get a frequency of $13.33 MHz$ into the audio codec from the $80 MHz$ frequency of Patmos.

\subsection{Audio I2C Component}

In the following the ${I}^{2}C$ communication with the audio chip is described. If there is a reference to the clock always the ${I}^{2}C$ clock is meant and not the clock from Patmos. 

The datasheet states that the maximum clock frequency is 526 kHz. It was decided to select 200 kHz as the clock frequency for \texttt{SCLK}, to be sure that there is enough time for the slave to read the data. 

${I}^{2}C$ requires a start condition which is the data transitioning from high to low while the clock stays high. After the start condition the slave address ( which is always 0011010) needs to be submitted concatenated with the read write signal. The write signal has to be always zero, because it is a write only device. After a successful delivery the device will answer with an acknowledge by pulling the data signal one clock cycle low. After that, the address for the configuration register (7-Bit) and the data (9-bit) can be submitted in two 8-bit chunks. The communication ends with a stop condition were the data signal goes from low to high while the clock is high.  

\begin{figure}[ht!]
\centering
\includegraphics[width=0.35\textwidth]{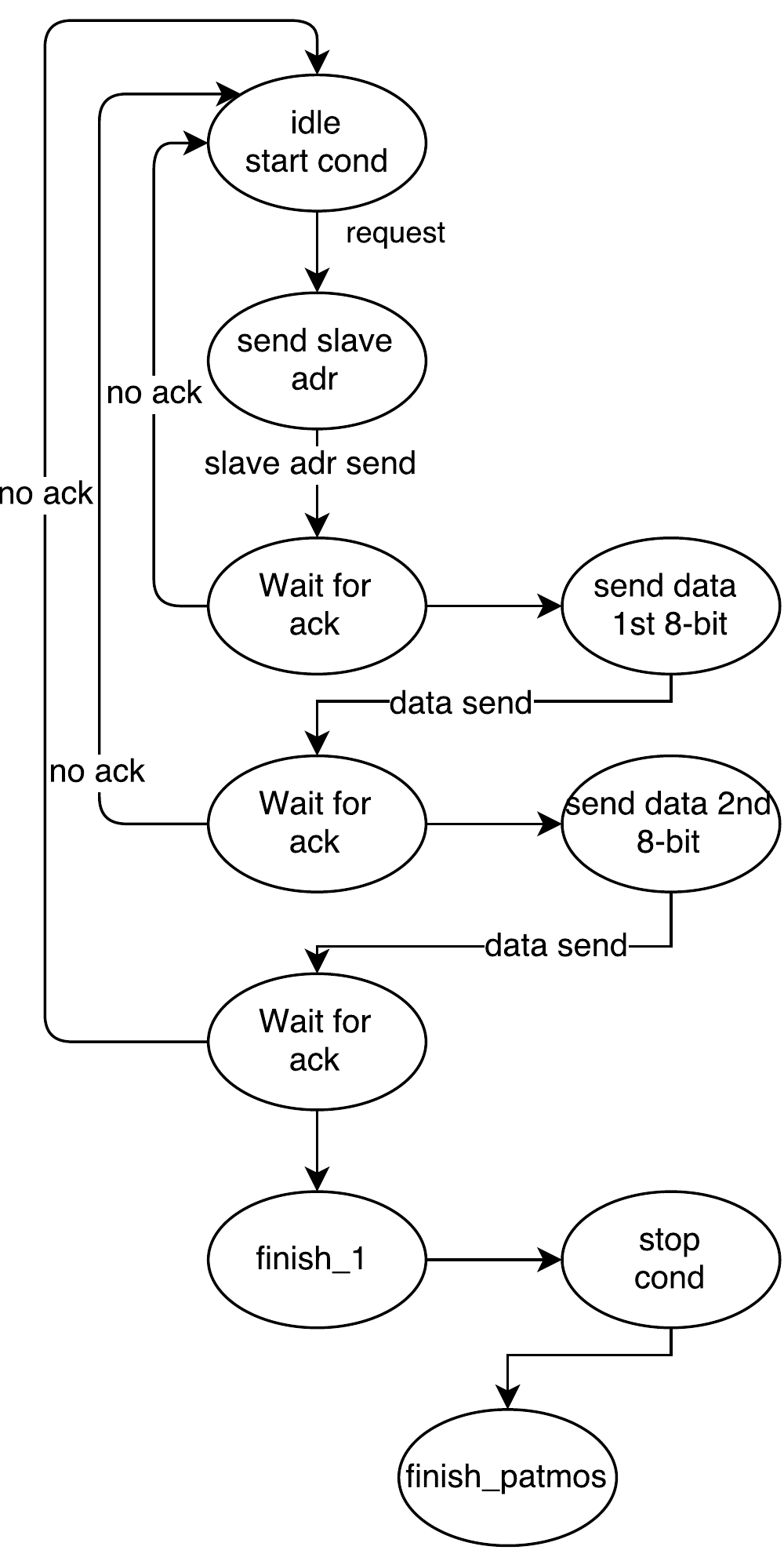}
\caption{State machine for the I2C communication}
\label{fig:SMI2C}
\end{figure}

Based on this setup configuration the sate machine shown in figure~\ref{fig:SMI2C} was created. 

\subsection{Audio ADC Component}

The "audio ADC" component is designed to receive audio data from \texttt{WM8731} and stores it into two audio registers for the left and right data. The bitlength of this registers is specified in the configuration XML file. In order to reduce the power consumption, an enable signal is used, so that the conversion is only done when Patmos requests so, setting the enable signal to high. While converting, the data should not be read, this is why the busy signal is used, so that Patmos makes sure it does not read any invalid data.

\begin{figure}[ht!]
\centering
\includegraphics[width=0.47\textwidth]{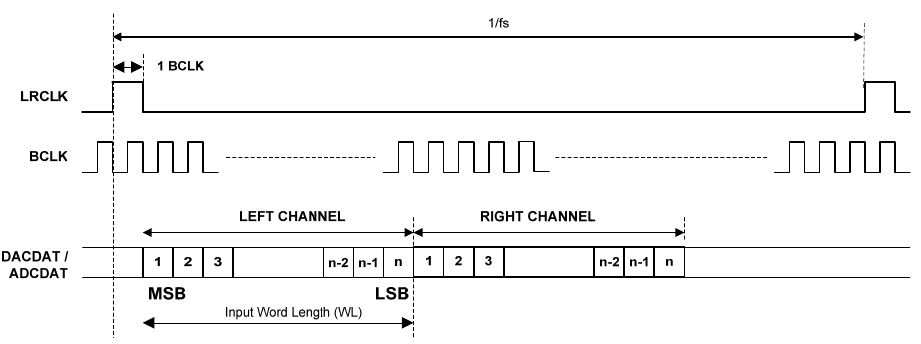}
\caption{DSP Mode A (source: \cite{WM8731DataSheet})}
\label{fig:DSPMode}
\end{figure}

For both the ADC and the DAC, the mode selected to receive or send audio data is the DSP Mode. As it can be seen from figure~\ref{fig:DSPMode} that first the LRC signal needs to be high for one \texttt{BCLK} clock cycle. Then, the audio codec will send the data of the left audio channel, and immediately after, the right channel. Furthermore, it can be seen that the most significant bit comes first. This communication mode is referred in the datasheet as the DSP mode A. 

The time between each LRC pulse has to fit the sampling frequency. The chosen sampling frequency is $48 KHz$, which the audio codec generates dividing its input clock, \texttt{XCLK} by 256. The \texttt{WM8731} audio codec expects an input clock frequency of $12.288 MHz$, to get $\frac{12.288*{10}^6}{256}=48 kHz$. However, this $12.288 MHz$ clock cannot be generated with an integer divider from the $80 MHz$ frequency of Patmos. One possible solution would be to change the clock of Patmos. Instead, the chosen solution has been to use a $13.33 MHz$ frequency of \texttt{BCLK}, so the resulting sampling frequency is $\frac{13.33*{10}^6}{256}=52.08 kHz$. This is not an standard sampling frequency value, but this is not extremely important here as the audio quality does not decrease (it actually increases slightly).

\begin{figure}[ht!]
\centering
\includegraphics[width=0.25\textwidth]{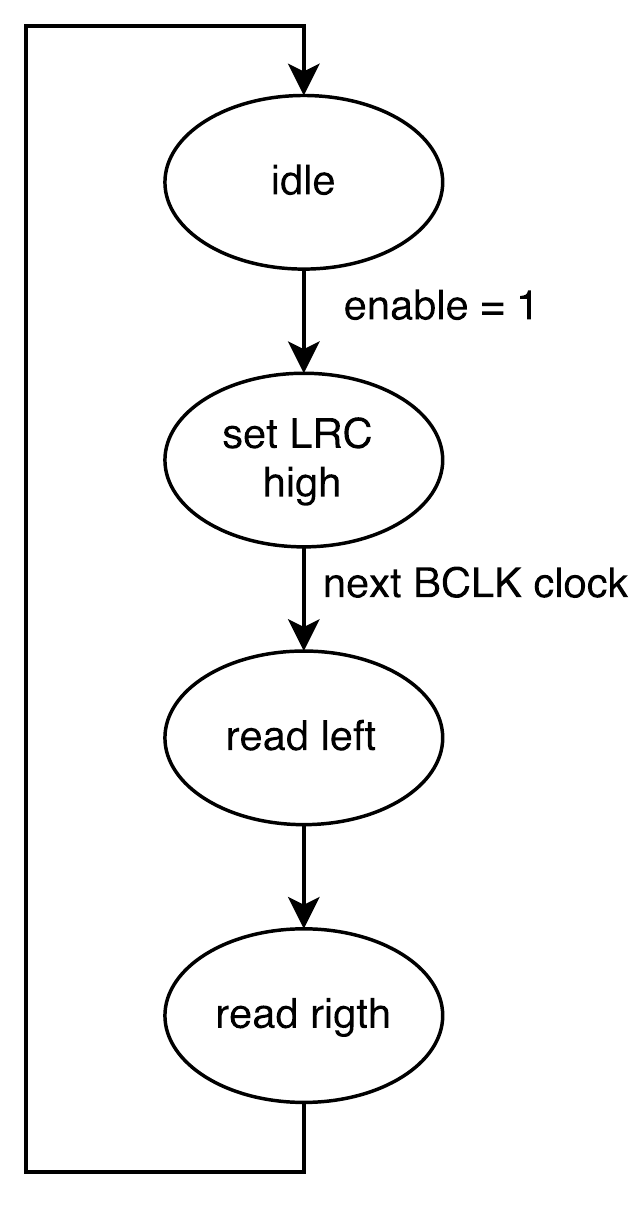}
\caption{State machine for the audio adc component}
\label{fig:SMADC}
\end{figure}

Based on this specification, the designed state machine for this module is shown in figure~\ref{fig:SMADC}. %Maybe some more text?

\subsection{Audio DAC Component}

The "audio DAC" component converts the output audio data, which is stored in registers inside the "audio interface" module, to the format need for the \texttt{WM8731}. The chosen conversion mode here is also DSP mode A, as in the "Audio ADC" component. Therefore, the design of the audio DAC is also very similar to the ADC described in the previous section. Instead of receiving audio data the data will be send to the audio chip. The state machine used to design this module in Chisel is not shown due to its similarity with the previous one. 

However, one of the main differences is that, in the "audio ADC", the data pin (\texttt{ADCDAT}) is an input (output from the audio codec), so it needs to be captured in the raising edge of \texttt{BCLK}. Instead, in the "audio DAC", the audio data pin (\texttt{DACDAT}) is an output (input to the audio codec): this means that the data has to be writen in the pin in the falling edge of \texttt{BCLK}, so that the audio chip can read the data on the raising edge.

\subsection{Audio Clock Generation Component}

The \texttt{WM8731} is set to slave mode, which means that it needs to receive the master clock (\texttt{XCLK}) and the digital audio bit clock (\texttt{BCLK}) from the FPGA. As it is stated in the datasheet (\cite[p42]{WM8731DataSheet}), the sampling rate of the audio device depends on the \texttt{XCLK} value chosen. As explained before, none of these frequencies could be matched, so a frequency of $13.33 MHz$ has been chosen to get a sample rate of $52.08 kHz$, which is equally valid. These parameters can be easily changed from the XML configuration file of Patmos and from the configuration registers of the \texttt{WM8731}.
\newpage
\section{Implementation}
\label{sec:Implementation}

This section briefly describes the implementation of the design in the Altera \texttt{DE2-115} board, and points out challenges and methods used to achieve a successful implementation. All the files are publicly available at:

\url{https://github.com/dsanz006/audio_patmos.git}

\subsection{I2C and ADC Implementation}
\begin{figure}[ht!]
\centering
\includegraphics[width=0.47\textwidth]{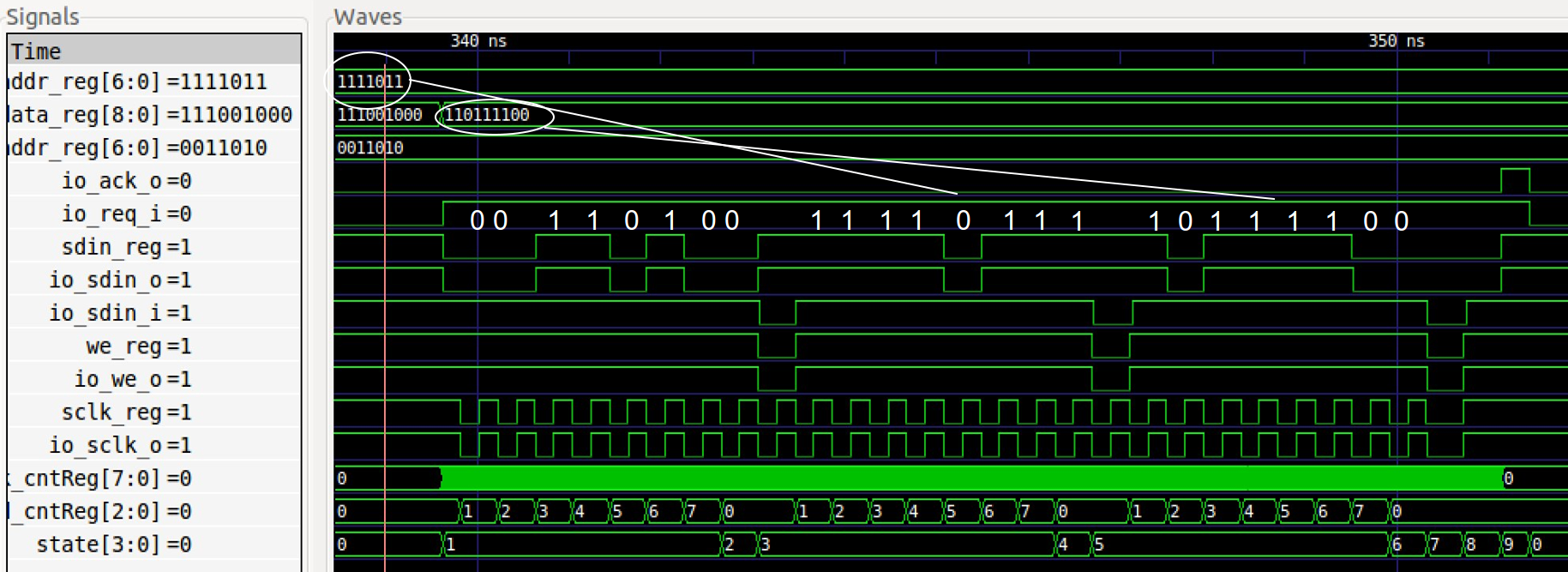}
\caption{Simulation of the I2C communication}
\label{fig:simi2c}
\end{figure}

The first goal was to get a minimal working solution and then extend the working design. For a minimal working solution it was both required to configure the registers and write something on the ADC channel. Initially, in order to verify that the modules mentioned before were implemented correctly, the Patmos emulator was used. Figure~\ref{fig:simi2c} shows the  ${I}^{2}C$ waveform graph. It can be seen that first interface address (0011010) plus the concatenate 0 are transmitted and then the address and data signals. Also the clock length is verified. After one 8-bit data is sent the \texttt{we} signal is set to low to let the audio codec pull \texttt{sdata} down as an acknowledge.

\begin{figure}[ht!]
\centering
\includegraphics[width=0.47\textwidth]{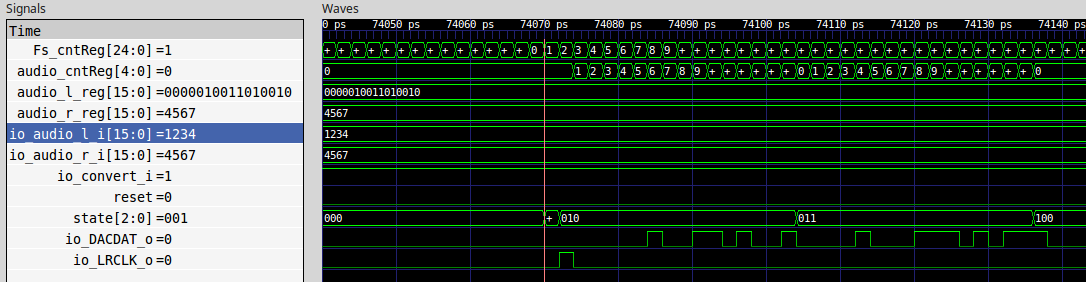}
\caption{Simulation of the DAC communication}
\label{fig:simAudioDAC}
\end{figure}

In figure~\ref{fig:simAudioDAC}, the simulation of DAC communication is shown. It can be seen that the LRC signal comes first, and then the left and right channel are transmitted. 

\subsection{SignalTap}

\begin{figure}[ht!]
\centering
\includegraphics[width=0.47\textwidth]{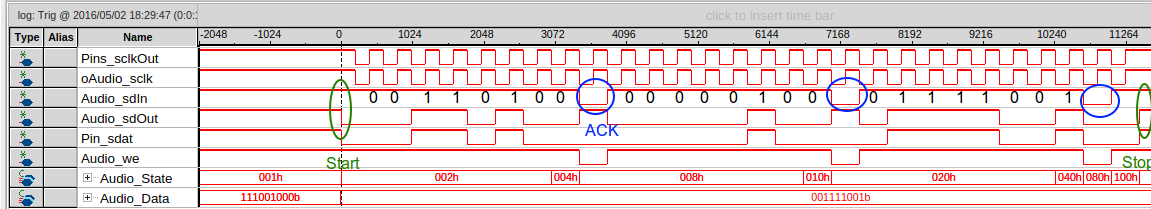}
\caption{SignalTap capture of the I2C communication}
\label{fig:SignalTapi2c}
\end{figure}

One of the main difficulties during the synthesis on the FPGA was that the configuration registers of the \texttt{WM8731} are read only: this means that it is not possible to check if data was written on them correctly.
With the help of the logic analyzer, SignalTap, that Quartus provides it was possible to get a better insight into what was happening. Looking at the waveforms from SingalTap helped immensely because it included the response of the audio chip. As it can be seen in figure~\ref{fig:SignalTapi2c} the acknowledge signal comes as desired after every eight bit submitted. This means that the data is being written into the configuration registers correctly. This can also also verified by checking some of the configuration parameters, such as the volume.

\begin{figure}[ht!]
\centering
\includegraphics[width=0.47\textwidth]{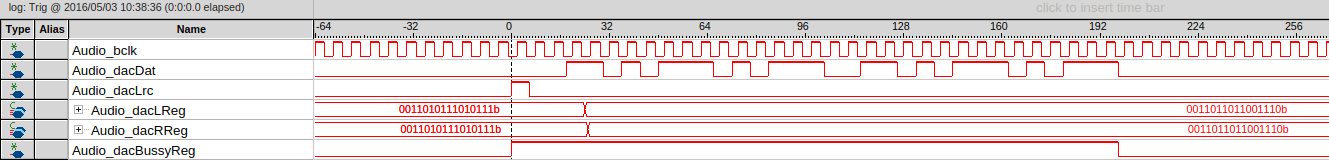}
\caption{SignalTap capture of the audio DAC communication in detail}
\label{fig:SignalTapDAC1}
\end{figure}

Again, SignalTap was used to check that the audio data was being sent and received correctly into and from the \texttt{WM8731}. Looking at the digital audio signals it could be verified that the format corresponds to DSP mode A, as expected. This is shown in figure~\ref{fig:SignalTapDAC1}. 

\begin{figure}[ht!]
\centering
\includegraphics[width=0.47\textwidth]{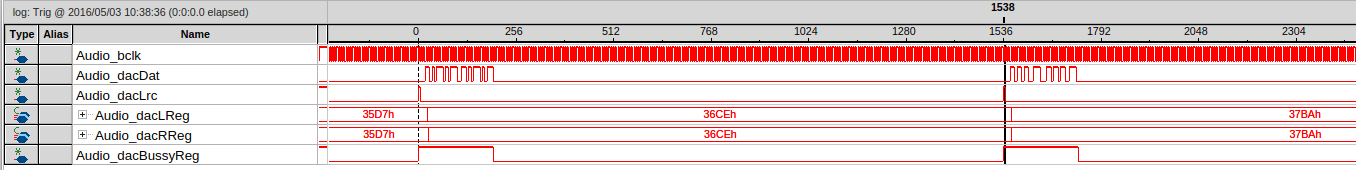}
\caption{SignalTap capture of the audio DAC communication overview}
\label{fig:SignalTapDAC2}
\end{figure}

In figure~\ref{fig:SignalTapDAC2} a submission of two consecutive audio packets can be seen. The \texttt{LRCK} signals are 1536 cycles apart. With a clock period of $80 MHz$, this corresponds to a period of $19.2 \mu s$ ($\frac{1536}{80*{10}^{6}}$). So the frequency is $\frac{1}{19.2*{10}^{-6}} = 52.08 kHz$. The logic analyzer was also used here to make the signals look exactly as it is specified in the datasheet. 

\subsection{C Implementation}

All the work done, using the Chisel hardware description language, is intended to make it easier to write programs to manipulate audio in many ways: recording and processing of a signal, synthesis, audio effects, and so on. It is much more suitable to write audio processing programs in a programming language such as C, and then run them on the Patmos processor instead of describing them in hardware. 

The C programs developed include the library (\texttt{audio.c}), where some functions can be found to set up the configuration registers, write and read audio to and from the audio codec and synchronise the audio signals. Several test programs have been written which verify that this functions work correctly. In a header file (\texttt{audio.h}) all the needed addresses for the available registers are defined.

The two most important functions implemented are the ones that read and write audio data from and to the \texttt{WM8731}. As a design decision, it was chosen not to create any audio buffer: this decision was chosen because Patmos is intended to be a real-time processor, and audio buffers introduce a delay from input to output, which can lead to a loss of the real-time concept. Instead, it has been decided that each audio sample is input, processed and output directly, with a delay of only 1 sample: $19.2 \mu s$. The way the audio input and output function works is very simple: data is simply written or read to or from the audio data registers in the audio interface (between Patmos and the \texttt{WM8731}). Then, the hardware module takes care of transmitting this signal to the \texttt{WM8731} only when the device is not busy, so when there is no conversion going on.

Additionally, it is necessary to synchronize each audio sample with the sampling frequency, to avoid overwriting the same values of data. Therefore, two synchronisation functions (for ADC and DAC) were created, which synchronize with the sampling frequency by waiting for the LRC pulse.

Finally, there is a function available to change the volume of the audio output. The volume can be set between +6 and -73 db. No functions have been implemented for other configuration registers, because the only parameter that is usually modified many times during execution is the volume.

\subsection{FPGA Tests}

A simple test created to validate the implementation is a program to output a sine wave. While this test was running, the analog output was recorded using an external audio interface, and it could be confirmed that the output looked exactly like a sinusoidal signal.

The second test was to output the data directly from the input, without modifying it. This was done using a program where the data from the input registers was copied into the output registers, and then waiting to the synchronization signal again. This program proved to work correctly: a drum machine was used to generate a drum loop audio signal, and it was recorded directly using the external audio interface. After that, the same audio signal was recorded again, but this time going into and out of the FPGA: both signals looked exactly the same, and also sounded the same. 

\begin{figure}[ht!]
\centering
\includegraphics[width=0.47\textwidth]{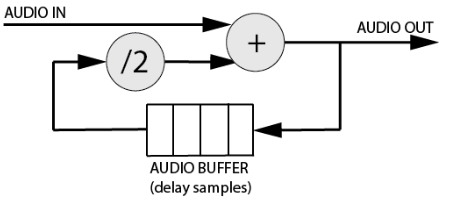}
\caption{Delay effect}
\label{fig:Delay}
\end{figure}

A more elaborated test was the implementation of a simple delay effect. The input audio was mixed with a delayed copy of input. The idea of the effect is illustrated in figure~\ref{fig:Delay}. Using the keys on the FPGA, the user had the possibility to manipulate the delay time from none to up to one second. Furthermore, it was possible to change the volume of the output. 

\section{Evaluation}
\label{sec:Evaluation}
% How did this project turn out.
% What could have been done differently
% Future work

The final results of this project show to be very satisfactory: the initial goal of designing an audio interface has been achieved, and it has been proved that it works correctly, both for inputting audio into Patmos, and for outputting it as well.

The workflow has been done sequentially, where each module has been simulated and tested in the FPGA before starting to design and implement new modules. As explained before, first the I2C module was implemented, then the DAC module, which was tested by synthesizing a sinusoidal signal, and finally the ADC module.

The C programs behave as expected, and the audio interface is flexible with the audio codec: it allows changing parameters easily, such as the sampling rate and the audio data bit length, as well as some other parameters such as the input and output volumes.

One of the main drawbacks of the design is the lack of an audio buffer between the input and the output signal, which is common in many computers, laptops, smartphones and many electronic devices nowadays. However, as explained before, this design decision was taken in order to reduce the audio latency as much as possible. But the implementation of an audio buffer is not difficult and can be considered as part of the future work.

\section{Conclusions}
\label{sec:conclusion}

As said in the previous section, the results of this project have proved to be satisfactory. An efficient audio interface has been designed which allows writing audio programs to Patmos.

We have learned a lot from different perspectives: first of all, we have understood how a real processor can handle different buses or ports which are connected to its IO pins, which is a pure advanced computer architecture concept. But, apart from that, we have also learned how an audio codec works, what the main parameters and options are, and how to treat audio signals, both in digital hardware and in software. 

Last but not least, it is important to discuss the future work that is related to this project. As mentioned before, the implementation of an audio buffer can be easily done: this would not only allow to relax the processor, but it would also permit the easier implementation of some very extended audio effects, such as the Fourier Frequency Transformation, where audio signals are not treated as a single sample, but as a block of samples instead. Apart from this, the use of the multi-core version of Patmos can be very useful to accelerate some audio computation for big programs. The digital hardware in the FPGA can also be used in the same way, for audio processing acceleration. This is something very common for video and graphics acceleration, but it can provide interesting improvements on audio computing as well.

\section{Acknowledgments}
This paper was written with in the scope of the course Advanced Computer Architecture (02211) at the Technical University of Denmark in the spring of 2016.

We would like to thank Martin Schoeberl (\texttt{masca@dtu.dk}) and Jens Spars{\o} (\texttt{jspa@dtu.dk}) for their support and supervision.
\newpage
% 
% The following two commands are all you need in the
% initial runs of your .tex file to
% produce the bibliography for the citations in your paper.
\bibliographystyle{abbrv}
\bibliography{sigproc}  % sigproc.bib is the name of the Bibliography in this case

\begin{thebibliography}{1}

\bibitem{thesis}
J.~I.~M. Albiach.
\newblock Interfacing a processor core in fpga to an audio system.
\newblock Master's thesis, Link\"oping University, 2006.

\bibitem{DE2115DataSheet}
Altera.
\newblock De2-115 user manual.
\newblock
  \url{ftp://ftp.altera.com/up/pub/Altera_Material/Boards/DE2-115/DE2_115_User_Manual.pdf}.
\newblock Date: 2016-04-19.

\bibitem{ChiselTut}
J.~Bachrach, K.~Asanovi\'{c}, and J.~Wawrzynek.
\newblock Chisel 2.2 tutorial.
\newblock Technical report, EECS Department, UC Berkeley, July 2015.

\bibitem{WM8731DataSheet}
W.~Microelectronics.
\newblock Wm8731 - portable internet audio codec with headphone driver and
  programmable sample rates.
\newblock \url{http://www.cirrus.com/en/pubs/proDatasheet/WM8731_v4.9.pdf}.
\newblock Date: 2016-04-19.

\bibitem{PatmosSpec}
M.~Schoeberl, F.~Brandner, S.~Hepp, W.~Puffitsch, and D.~Prokesche.
\newblock Patmos reference handbook.
\newblock Technical report, DTU Compute, Technical University of Denmark,
  October 2015.

\end{thebibliography}
% You must have a proper ".bib" file
%  and remember to run:
% latex bibtex latex latex
% to resolve all references
%
% ACM needs 'a single self-contained file'!
%
%APPENDICES are optional
%\balancecolumns

% This next section command marks the start of
% Appendix B, and does not continue the present hierarchy
\balancecolumns
% That's all folks!
\end{document}